# Quantum Magnetic Excitations from Stripes in Copper-Oxide Superconductors

J. M. Tranquada[1], H. Woo[1,2], T. G. Perring[2], H. Goka[3], G. D. Gu[1], G. Xu[1], M. Fujita[3], and K. Yamada[3]

*[1]Physics Department, Brookhaven National Laboratory, Upton, NY 11973, USA*
*[2]ISIS Facility, Rutherford Appleton Laboratory, Chilton, Didcot, Oxon OX11 0QX, UK*
*[3]Institute for Materials Research, Tohoku University, Sendai, 980-8577, Japan*

**In the copper-oxide parent compounds of the high-transition-temperature superconductors,[1] the valence electrons are localized, one per copper site, due to strong intraatomic Coulomb repulsion. A symptom of the localization is antiferromagnetism,[2] where the spins of localized electrons alternate between up and down. The superconductivity appears when mobile 'holes' are doped into this insulating state, and it coexists with antiferromagnetic fluctuations.[3] In one approach to the coexistence, the holes are believed to self-organize into 'stripes' that alternate with antiferromagnetic (insulating) regions within copper-oxide planes.[4] Such an unusual electronic state would necessitate an unconventional mechanism of superconductivity.[5] There is an apparent problem with this picture, however: measurements of magnetic excitations in superconducting $YBa_2Cu_3O_{6+x}$ near optimum doping[6] are incompatible with the naive expectations[7,8] for a material with stripes. Here we report neutron scattering measurements on stripe-ordered $La_{1.875}Ba_{0.125}CuO_4$. We show that the measured excitations are, surprisingly, quite similar to those in $YBa_2Cu_3O_{6+x}$[9,10] (i.e., the predicted spectrum of magnetic excitations[7,8] is wrong). We find instead that the observed spectrum can be understood within a stripe model by taking account of *quantum* excitations. Our results support the concept that stripe correlations are essential to high-transition-temperature superconductivity.[11]**

La$_{2-x}$Ba$_x$CuO$_4$ ("Zurich" oxide) is the material in which Bednorz and Müller[1] first discovered high-temperature superconductivity. An anomalous suppression of the superconductivity found[12] in a very narrow region about $x = 1/8$ was later shown to be associated with charge and spin stripe order.[4,13] Schematic diagrams of stripe order in the CuO$_2$ planes are shown in Fig. 1(a) and (b). Between the charge stripes are regions with locally-antiferromagnetic order. The "bond-centered" stripe model shown has received considerable theoretical attention.[11,14-16]

In a uniform (undoped), two-dimensional, antiferromagnetic CuO$_2$ plane, with lattice parameter $a = 3.8$ Å between the nearest-neighbor Cu ions, a neutron diffraction measurement yields superlattice peaks characterized by the wave vector $\mathbf{Q}_{AF} = (1/2, 1/2)$, in units of $2\pi/a$. When vertical stripe order is present, the superlattice peaks move to $(1/2 \pm \delta, 1/2)$, where $\delta = 1/(2p)$, and $p$ is the charge stripe spacing; $p = 4$ in the present case. Analogous peaks appear for the case of horizontal stripes [see Fig. 1(f),(g)]. In $La_{1.875}Ba_{0.125}CuO_4$, both stripe orientations are present, so one observes both sets of superlattice peaks simultaneously [see Fig. 1(i)].[13]

The ordered magnetic moments can fluctuate about their average orientations. It costs energy to create these fluctuations, and the variation of the magnetic excitation energy with wave vector $\mathbf{Q}$ can be measured by inelastic neutron scattering. Antiferromagnetic $La_2CuO_4$ exhibits conventional spin-wave behavior,[17,18] with an energy dispersion $\hbar\omega = c|\mathbf{Q} - \mathbf{Q}_{AF}|$ for $\hbar\omega < J$, with $J$ ($\approx 140$ meV)[17] being the effective superexchange coupling between neighboring magnetic moments. The spin-wave velocity $c$ is proportional to $J$, and the spin waves extend up to a maximum energy of $2J$. The strength of the magnetic scattering is relatively large throughout the antiferromagnetic Brillouin zone centered on $\mathbf{Q}_{AF}$ [indicated by the dashed-line diamond in Fig. 1(f),(g),(i)], but becomes negligible outside of that zone.

One might expect to find spin waves dispersing out of the superlattice peaks of $La_{1.875}Ba_{0.125}CuO_4$,[7,8] in the form of expanding cones as shown in Fig. 1(d). The dispersion of spin waves has recently been measured in the stripe-ordered model compound $La_{2-x}Sr_xNiO_4$,[19,20] where $S = 1/2$ Cu spins are replaced by $S = 1$ Ni spins and the stripes run diagonally. In $La_{1.875}Ba_{0.125}CuO_4$ (Ref. 13) and a related compound,[21] the low-energy excitations (< 12 meV) behave similarly to spin waves. To find out what happens at higher energies, we have grown large, high-quality crystals of $La_{1.875}Ba_{0.125}CuO_4$ and studied them with a powerful time-of-flight neutron spectrometer, MAPS, at the ISIS spallation source.

Experimental measurements are presented in Fig. 2. These are constant-energy slices through though the magnetic scattering, with the energy transfer, $E = \hbar\omega$, increasing from the lower left to upper right. The data are plotted within an antiferromagnetic Brillouin zone, making use of the rotated coordinate system indicated at the bottom of Fig. 1. The magnetic scattering is centered on $\mathbf{Q'}_{AF}$ (where the ' denotes wave vectors in the rotated system). At 6 meV, one can clearly see the four incommensurate peaks, corresponding to fluctuations about the ordered stripe state. These low-energy results are similar to what is observed in superconducting $La_{2-x}Sr_xCuO_4$.[22] By 36 meV, the signal has dispersed inwards towards $\mathbf{Q'}_{AF}$, nearly forming a ring around that point. A simple commensurate peak is found at 55 meV.

Using a spin-wave model, we would expect the 36-meV data to look something like Fig. 3(a): four rings centered on the incommensurate wave vectors, representing cuts through the spin-wave dispersion cones. This picture is clearly quite different from the experimental results. The behavior at higher energies is even more striking. At 105 meV, the excitations have started to disperse outwards, but note the new shape: it is roughly a diamond, with points rotated 45° from the incommensurate wave vectors. This diamond continues to grow at 160 and 200 meV. Overall, the evolution of the magnetic scattering with energy looks amazingly similar to that measured by Hayden *et al*.[9] in superconducting $YBa_2Cu_3O_{6.6}$ (see Fig. 2 in Ref. 9).

The dispersion measured along $\mathbf{Q'} = (1+q, q, 0)$ is presented in Fig. 4(b).  Another interesting quantity to consider is the function $S(\omega)$, obtained by integrating the magnetic scattering intensity $S(\mathbf{Q},\omega)$ over $\mathbf{Q}$.  The results are shown in Fig. 4(a).  With increasing energy, $S(\omega)$ initially decreases, and then rises to a broad peak near 50–60 meV.  At higher energies, $S(\omega)$ gradually decreases.  These results are qualitatively similar to earlier results on La$_{2-x}$Sr$_x$CuO$_4$.[24]

One generally determines the nature of magnetic fluctuations from the ordered state with which they are associated.  In the case of La$_2$CuO$_4$, the high-energy spin waves are clearly associated with the antiferromagnetic order.   While the excitations in our sample are clearly different from semiclassical spin waves, we nevertheless expect them to be associated with the stripe order indicated by magnetic and charge-order superlattice peaks.[13]

Is there a simple way to interpret our observations? If, for the moment, we ignore the low-energy incommensurate scattering, the finite-energy peak in $S(\omega)$ suggests that we are measuring singlet-triplet excitations of decoupled spin clusters.  Given the stripe order in our sample, an obvious candidate for such a cluster would be one of the magnetic domains shown in Fig. 1(a) and (b), corresponding  to what is commonly called a 2-leg spin ladder [Fig. 1(c)].  (This name refers to the pattern formed by the exchange paths between the magnetic ions.)  Now, a spin ladder has rather interesting properties.[25]  The superexchange $J$ between neighboring spins keeps them antiparallel, but there is no static order at any temperature.  This fluctuating, correlated state is said to be quantum disordered.  There is a substantial energy gap ($\approx 0.5J$) to the first excited state, and the excitations disperse only along the ladder direction, not along the rungs [see Fig. 1(e)].

To compare with experiment, we have calculated simulated spectra, Fig. 3 (b)-(e), using the single-mode approximation for the scattering function of a spin ladder with isotropic exchange (I. Zaliznyak, unpublished)

$$S(\mathbf{Q},\omega) \sim (\hbar\omega_{q\parallel})^{-1} [\sin^2(q_\parallel a/2) + \sin^2(q_\perp a/2)] [\delta(\omega - \omega_{q\parallel}) - \delta(\omega + \omega_{q\parallel})].$$

Here, $q_\parallel$ is measured parallel to the ladder, $q_\perp$ is along the rungs, and the dispersion $\omega_{q\parallel}$, which is proportional to $J$, is given by Barnes and Riera[26] [see Fig. 1(e)].  Parts of this scattering function have been tested in measurements of ladder excitations on Sr$_{14}$Cu$_{24}$O$_{41}$.[27]  In the simulations, we see that the most intense signal has a diamond shape that disperses outward with energy, similar to the right-hand side of Fig. 2.

The calculated and measured $S(\omega)$ and $\omega(q)$ are compared in Fig. 4 (a) and (b), respectively.  The agreement is remarkable considering the simplicity of the model.  The energy scale for the dispersion is set by a single parameter, $J$, and the value of $J$ is only modestly reduced from that in the parent compound, La$_2$CuO$_4$.  The downward dispersion below 50 meV can be modelled by allowing weak coupling between the ladders, through the charge stripes, as demonstrated by unpublished simulations of R. Konik and F. Essler.

For completeness, we note that dispersions with similarities to our data have also been obtained in weak-coupling, itinerant-electron calculations.[28,29] While this approach provides a possible alternative explanation of our results, one should be aware that explaining the observed energy scale for the excitations would require fine tuning of parameters,[4] and that explaining the charge-order in our sample is an unmet challenge for the weak-coupling approach. Thus, we believe that the ladder model, within the stripe picture, provides a more compelling explanation of the results. Given the similarity with recent measurements[9,10] on $YBa_2Cu_3O_{6+x}$, together with the evidence for spatially-anisotropic magnetic excitations in detwinned $YBa_2Cu_3O_{6+x}$,[30] our results provide support for the concept that charge inhomogeneity, possibly dynamic in nature, is essential to achieve superconductivity with a high transition temperature in copper oxides.[5,11]

**Acknowledgements**  JMT, HW, GDG and GX are supported by the Office of Science, U.S. Department of Energy.  KY and MF are supported by the Japanese Ministry of Education, Culture, Sports, Science and Technology. Work supported in part by the U.S.–Japan Cooperative Research Program on Neutron Scattering.  We gratefully acknowledge assistance with sample characterization from A. R. Moodenbaugh and Qiang Li, and helpful discussions with E. Carlson, F. Essler, S. A. Kivelson, R. Konik, S. Sachdev, and I. Zaliznyak.

**Competing interests statement**  The authors declare that they have no competing financial interests.

**Correspondence** and requests for materials should be addressed to JMT (e-mail: jtran@bnl.gov).

**Figure 1** Schematic diagrams illustrating stripe order in real space and possible magnetic scattering in reciprocal space. (a) Real-space diagram of vertical stripes; (b) horizontal stripes; (c) a horizontal spin ladder; circles indicate Cu sites in hole-doped stripes, arrows indicate magnetic moments on undoped Cu sites. (While the specific alignment of the stripes with respect to the lattice has not been directly determined by experiment, the "bond-centered" stripes shown here suggest a natural model for interpreting our results.)  (d) Dispersion of spin waves due to vertical stripes; (e) triplet excitations in a horizontal spin ladder. [For the latter, the minimum spin gap appears along $\mathbf{Q}$ = (1/2, $k$), for all $k$, and the maximum energy is reached along $\mathbf{Q}_m$ = (1/2±1/4, $k$).[26]  Because of the singlet correlations, the magnetic scattering is relatively strong for $\mathbf{Q}$ between the lines $\mathbf{Q}_m$, but weak outside of them.  Thus, these lines define a zone of strong magnetic intensity.]  (f) Reciprocal space, with $h$ and $k$ in reciprocal lattice units (rlu), showing incommensurate magnetic wave vectors for vertical stripe order (dashed line gives antiferromagnetic zone boundary); (g) same for horizontal stripes; (h) intensity zone boundary for a horizontal spin ladder. (i) Wave vectors and boundaries expected when both orientations of stripes [and (j) ladders] are present simultaneously. (k) Rotation of the coordinate system by 45° changes the antiferromagnetic zone from a diamond into a square box, the latter being more convenient for plotting the data. (l) Incommensurate points in the rotated system, where $\mathbf{Q'}_{AF}$ = (1,0,0) measured in units of $2\pi/(a\sqrt{2})$. (We will use ' to indicate wave vectors in the rotated system.)  (m) Expanded view of a single AF zone in the rotated system; (n) same showing the zone of strong magnetic scattering for ladders, which forms a diamond.

**Figure 2** Constant-energy slices through the experimentally-measured magnetic scattering from $La_{1.875}Ba_{0.125}CuO_4$. Intensity, measured at T = 12 K (> $T_c$ ), is plotted in false color within a single antiferromagnetic zone [c.f. Fig. 1(m),(n)].  Energy has been integrated over the ranges indicated by the error

bars, and Q dependence has been convolved with a Gaussian to reduce scatter.  Panels (a)-(c) measured with an incident neutron energy $E_i$ = 80 meV, (d) -(g) with $E_i$ = 240 meV, and (h) with $E_i$ = 500 meV.   (In addition to the magnetic scattering, there are also features due to phonons, and "background" from single- and multiple-phonon scattering.)  The sample consisted of four crystals, with a total mass of 58 g, grown by the travelling-solvent floating-zone method at Brookhaven.  Magnetic susceptibility measurements on pieces cut from the ends of each crystal show that $T_c$ is generally less than 3 K, but at one end of each of two crystals it is close to 6 K.  The crystal quality was first checked on the BT-9 spectrometer at the National Institute for Science and Technology (NIST) Center for Neutron Research in Gaithersburg, MD, while the mounting and alignment were performed on the KSD spectrometer at the JRR-3M reactor in Tokai, Japan.  For the experiment at the MAPS spectrometer, the co-aligned crystals were oriented with their $c$-axes parallel to the incident beam.

**Figure 3** Simulations of constant-energy slices.  (a) Spin-wave model for $\hbar\omega$ = 36 meV.  (b)-(g) 2-leg ladder model, discussed in the text, using $J$ = 100 meV.  For the latter, the $\delta$-functions in frequency are replaced by Lorentzians with energy width $\Gamma$ = 0.2$J$.  In all cases, we have averaged over both orientations (horizontal and vertical) of stripes or spin ladders, and have integrated over the same energy ranges as in Fig. 2.

**Figure 4** Experimental results for integrated magnetic scattering and dispersion of the excitations.  (a) $S(\omega)$, as defined in the text.  Circles: $E_i$ = 80 meV data set; squares: $E_i$ = 240 meV; diamonds: $E_i$ = 500 meV.  In distinguishing the magnetic scattering from other signals, care was taken to avoid strong contributions from phonon branches at 20 and 47 meV. To obtain just the spin-dependent behavior, we have corrected for the anisotropic magnetic form factor.[23]  Further investigation is required to determine whether or not the sharp feature at 42 meV is actually magnetic.  (b) Dispersion measured along **Q'** = (1+$q$, $q$), with the assumption of symmetry about $q$ = 0. Red lines in (a) and (b) are calculated from the 2-leg spin ladder model with the same parameters as in Fig. 3.  Black dashed line in (a) is a Lorentzian to describe the low energy signal, and the red dot-dashed line is the sum of the other two curves.  Vertical "error" bars in (a) and (b) indicate the energy range over which data were integrated, while horizontal bars in (b) indicate the half-widths in $q$ of the fitted Gaussian peaks.

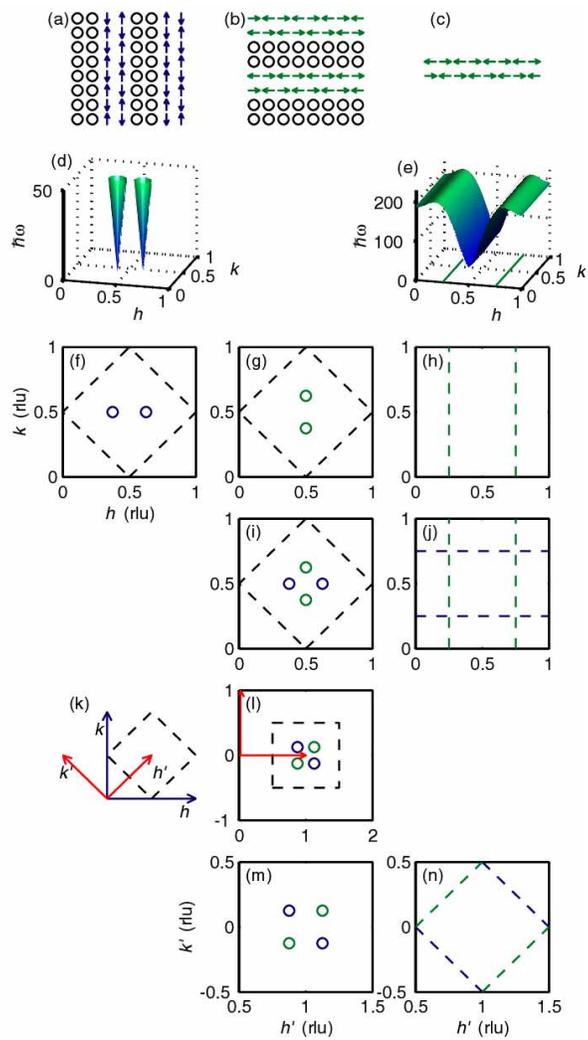

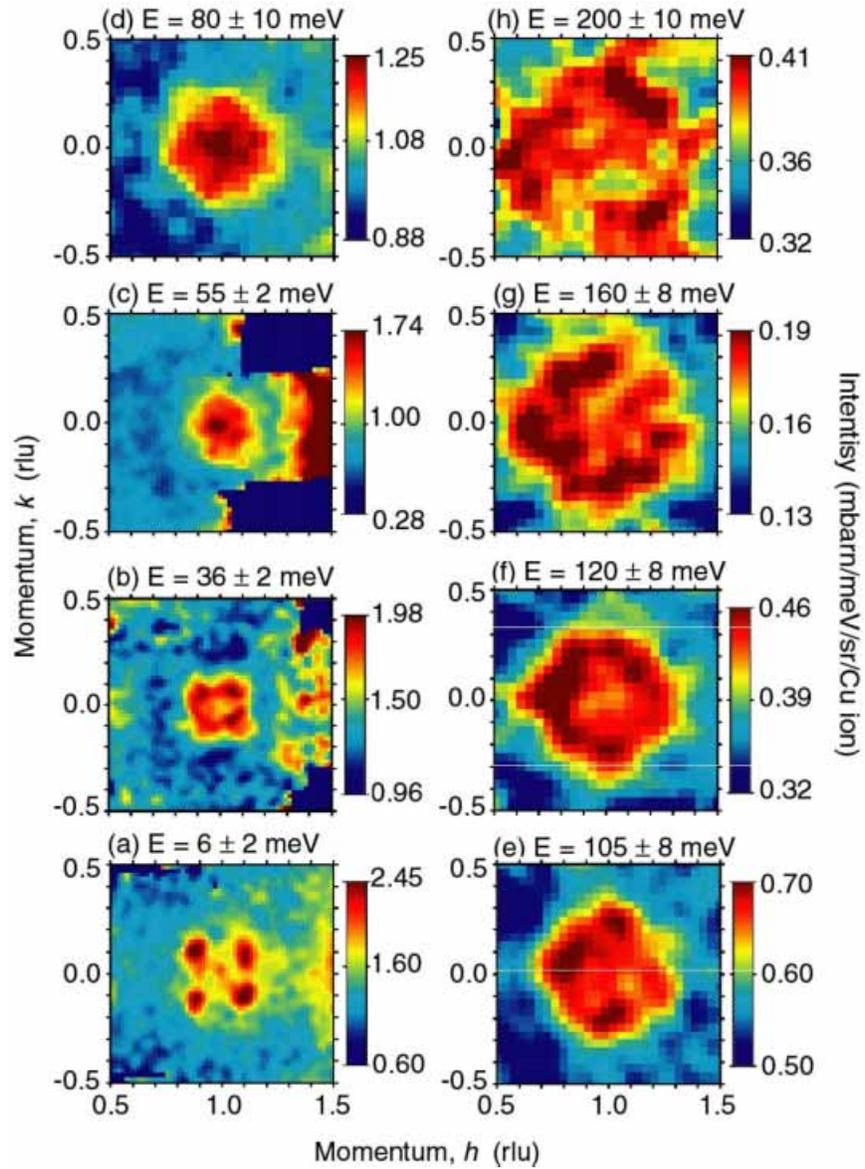

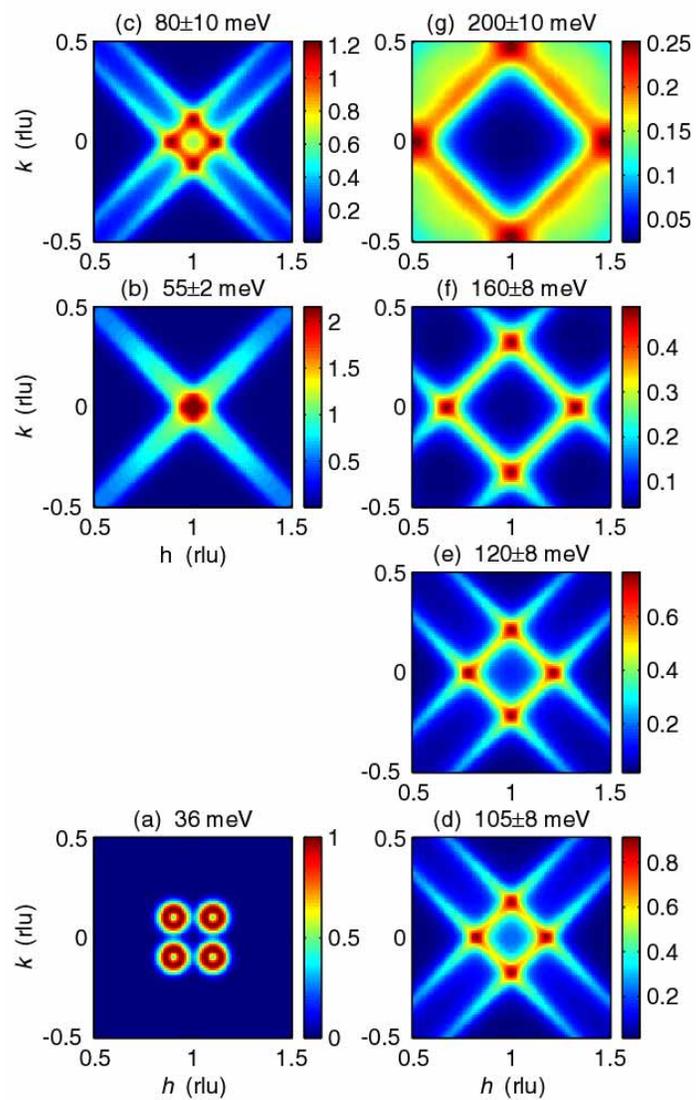

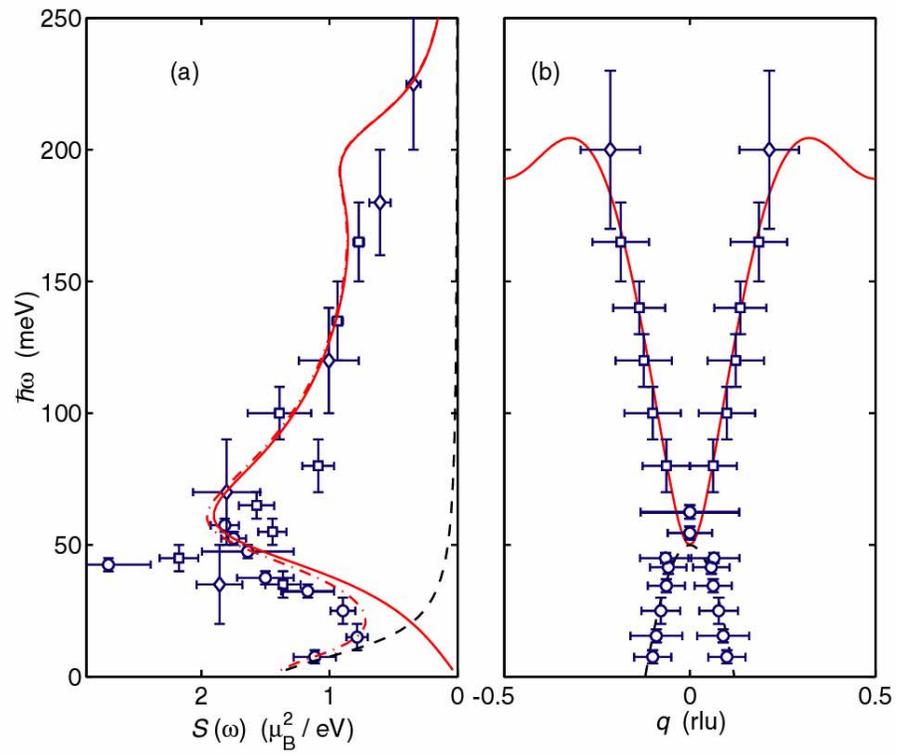